\documentclass[12pt]{article}
\usepackage[dvips]{color}
\usepackage{epsfig}
\usepackage{amsmath}
\usepackage{graphicx}
\def\Box{\hbox{$\rlap{$\sqcup$}\sqcap$}}
\begin{document}
\title{\bf Formulation of the Spinor Field in the Presence of a Minimal Length Based on the Quesne-Tkachuk Algebra}

\author{S. K. Moayedi $^{a}$\thanks{E-mail: s-moayedi@araku.ac.ir}\hspace{1mm}
, M. R. Setare $^{b}$ \thanks{E-mail: rezakord@ipm.ir}\hspace{1mm}
and
 H. Moayeri $^{a}$\thanks{E-mail: h-moayeri@phd.araku.ac.ir}\hspace{1mm} \\
$^a$ {\small {\em  Department of Physics, Faculty of Sciences,
Arak University, Arak 38156-8-8349, Iran}}\\
$^{b}${\small {\em Department of Science, Sanandaj Branch,
Islamic Azad University, Sanandaj, Iran
}}\\
}
\date{\small{}}
\maketitle
\begin{abstract}
\noindent

In 2006 Quesne and Tkachuk (J. Phys. A: Math. Gen. {\bf 39},
10909, 2006) introduced a (D+1)-dimensional
$(\beta,\beta')$-two-parameter Lorentz-covariant deformed algebra
which leads to a nonzero minimal length. In this work, the
Lagrangian formulation of the spinor field in a (3+1)-dimensional
space-time described by Quesne-Tkachuk Lorentz-covariant deformed
algebra is studied in the case where $\beta'=2\beta$ up to first
order over deformation parameter $\beta$. It is shown that the
modified Dirac equation which contains higher order derivative of
the wave function describes two massive particles with different
masses. We show that physically acceptable mass states can only
exist for $\beta<\frac{1}{8m^{2}c^{2}}$. Applying the condition
$\beta<\frac{1}{8m^{2}c^{2}}$ to an electron, the upper bound for
the isotropic minimal length becomes about $3 \times 10^{-13}m$.
This value is near to the reduced Compton wavelength of the
electron $(\mathchar'26\mkern-10mu\lambda_{c}=
\frac{\hbar}{m_{e}c}=3.86\times 10^{-13} m)$ and is not
incompatible with the results obtained for the
minimal length in previous investigations. \\

\noindent
\hspace{0.35cm}

{\bf Keywords:} Quantum gravity; Generalized uncertainty
principle; Minimal length; Spinor field; Dirac equation

\end{abstract}
\section{Introduction}
The unification between general relativity and quantum mechanics
is one of the major subjects of recent studies in theoretical
physics. The most interesting consequence of the above
unification is that in quantum gravity there is a minimal
observable distance on the order of the Planck length,
$l_{P}=\sqrt{\frac{G\hbar}{c^{3}}}\simeq 1.6 \times 10^{-35}m$,
where $G$ is the Newton's constant. The existence of this minimal
observable length which is motivated from various theories of
quantum gravity (such as perturbative string theory) and black
hole gedanken experiments, leads to a generalization of Heisenberg
uncertainty principle. This generalized or gravitational
uncertainty principle (GUP) can be written as
\begin{equation}
\triangle X\geq \frac{\hbar}{2\triangle
P}+\frac{\hbar}{2}\beta\triangle P,
\end{equation}
where $\beta$ is a positive parameter [1-19]. At high energies,
the second term on the right-hand side of (1) becomes significant
and leads to important deviations from the usual quantum
mechanics. With the GUP even at high momenta $\triangle X$ is
limited in resolution because of quantum gravitational effects.
In other words, independent of momentum, $\triangle X$ is always
larger than a minimal observable length $(\triangle
X)_{min}=\hbar\sqrt{\beta}.$ Nowadays, physicists are trying to
reformulate the quantum field theory in the presence of a minimal
observable length and there is hoping that this approach causes
unwanted divergencies can be eliminated or modified in quantum
field theory [7]. In Ref. [10], the real Klein-Gordon field in
the presence of a minimal observable length was studied and the
authors estimated the minimal observable length must be in the
range $10^{-17}m<(\triangle X)_{min}<10^{-15}m$. SI units are used throughout this work.\\
This paper is organized as follows. In Sect. 2, the
(3+1)-dimensional $(\beta,\beta')$-two-parameter Lorentz-covariant
deformed algebra introduced by Quesne and Tkachuk is reviewed and
it is shown that the above algebra leads to a minimal observable
length [5,6]. In Sect. 3, the Lagrangian formulation of the Dirac
spinor field in a (3+1)-dimensional space-time described by
Quesne-Tkachuk algebra is presented in the case where
$\beta'=2\beta$ up to first order over deformation parameter
$\beta$. In Sect. 4, the solutions of the modified spinor field
equation for free motion of a Dirac particle are obtained and it
is shown that these solutions are associated with two different
mass states. We find that the minimal observable length in the
modified spinor theory is of the order of about $3 \times
10^{-13}m$. Finally, Sect. 5 is devoted to the conclusions.

\section{A Brief Review of the Quesne-Tkachuk Algebra}
Let us start with a quick review of the Quesne-Tkachuk algebra,
which is a Lorentz-covariant deformed algebra that describes a
$(D+1)$-dimensional quantized space-time [5,6]. The
$(3+1)$-dimensional Quesne-Tkachuk algebra is characterized by
the following modified commutation relations
\begin{equation}
[X^{\mu},P^{\nu}]=-i\hbar(g^{\mu\nu}(1-\beta
P_{\rho}P^{\rho})-\beta'P^{\mu}P^{\nu}),
\end{equation}
\begin{equation}
[X^{\mu},X^{\nu}]=i\hbar\frac{2\beta-\beta'-(2\beta+\beta')\beta
P_{\rho}P^{\rho}}{1-\beta
P_{\rho}P^{\rho}}(P^{\mu}X^{\nu}-P^{\nu}X^{\mu}),
\end{equation}
\begin{equation}
[P^{\mu},P^{\nu}]=0,
\end{equation}
where $\mu,\nu,\rho =0,1,2,3$ and $\beta,\beta'$ are two
deformation parameters which are assumed non-negative
$(\beta,\beta'\geq 0)$. In terms of length $(L)$, mass $(M)$, and
time $(T)$ the deformation parameters $\beta$ and $\beta'$ have
the same dimensions $M^{-2}L^{-2}T^{2}$, i.e.,
$[\beta]=[\beta']=(momentum)^{-2}$. Also, $X^{\mu}$ and $P^{\mu}$
are deformed position and momentum operators and
$g_{\mu\nu}=g^{\mu\nu}=diag(1,-1,-1,-1)$. Using (2) and the
Schwarz inequality for a quantum state, the uncertainty relation
for position and momentum by assuming that $\triangle P^{i}$ is
isotropic $(\triangle P^{i}=\triangle P,\hspace{2mm} i=1,2,3)$
becomes
\begin{equation}
\triangle X^{i}\triangle
P\geq\frac{\hbar}{2}\left|1-\beta\left\{\langle(P^{0})^{2}\rangle
-3(\triangle P)^{2}-\sum^{3}_{j=1}\langle
P^{j}\rangle^{2}\right\}+\beta'\left[(\triangle P)^{2}+\langle
P^{i}\rangle^2\right]\right|.
\end{equation}
Hence, we arrive at an isotropic absolutely smallest uncertainty
in position given by
\begin{equation}
(\triangle
X^{i})_{0}=\hbar\sqrt{(3\beta+\beta')\left[1-\beta\langle
(P^{0})^{2}\rangle\right]}\hspace{2mm}, \hspace{5mm}i\in
\{1,2,3\}.
\end{equation}
In [8,11], Samar and Tkachuk introduced a representation which
satisfies the modified commutation relations (2)-(4) up to first
order in $\beta,\beta'$. The Samar-Tkachuk representation is
given by
\begin{equation}
X^{\mu}=x^{\mu}- \frac{2\beta -
\beta'}{4}(x^{\mu}p^{2}+p^{2}x^{\mu}),
\end{equation}
\begin{equation}
P^{\mu}=(1-\frac{\beta'}{2}p^{2})p^{\mu},
\end{equation}
where $x^{\mu}, p^{\mu}=i\hbar\frac{\partial}{\partial
x_{\mu}}=i\hbar\partial^{\mu}$ are position and momentum
operators in ordinary relativistic quantum mechanics, and
$p^{2}=p_{\alpha}p^{\alpha}=(p^{0})^{2}-\bf{p}.\bf{p}$ .\\
In this paper, we only consider the special case $\beta'=2\beta$,
wherein the position operators commute with each other in linear
approximation over the deformation parameters, i.e.,
$[X^{\mu},X^{\nu}]=0$. In such a linear approximation, the
Quesne-Tkachuk algebra reads
\begin{equation}
[X^{\mu},P^{\nu}]=-i\hbar(g^{\mu\nu}(1-\beta
P_{\rho}P^{\rho})-2\beta P^{\mu}P^{\nu}),
\end{equation}
\begin{equation}
[X^{\mu},X^{\nu}]=0,
\end{equation}
\begin{equation}
[P^{\mu},P^{\nu}]=0.
\end{equation}
It is easy to show that the following representations satisfy
(9)-(11), at the first order in $\beta$,
\begin{equation}
X^{\mu}=x^{\mu},
\end{equation}
\begin{equation}
P^{\mu}= (1-\beta p^{2})p^{\mu}.
\end{equation}
It should be noted that the representations (7),(8) and (12),(13)
coincide when $\beta'=2\beta$.

\section{Lagrangian Formulation of the Spinor Field Based on the Quesne-Tkachuk Algebra}
The Dirac Lagrangian density for a spinor $(spin-\frac{1}{2})$
field is [20]
\begin{equation}
{\cal L}(\Psi , \overline{\Psi} , \partial_{\mu}\Psi ,
\partial_{\mu}\overline{\Psi})= \frac{i\hbar c}{2}\left[\hspace{1mm}\overline{\Psi}\gamma^{\mu}(\partial_
{\mu}\Psi)-(\partial_{\mu}\overline{\Psi})\gamma^{\mu}\Psi\right]-mc^{2}\overline{\Psi}\Psi,
\end{equation}
where $\Psi$ is a Dirac spinor, $\gamma^{\mu}$ are the Dirac
matrices, and $\overline{\Psi} :=\Psi^{\dag} \gamma^{0}$ is the
adjoint spinor. The Euler-Lagrange equation for $\overline{\Psi}$
is
\begin{equation}
\frac{\partial {\cal L}}{\partial
\overline{\Psi}}-\partial_{\mu}\left(\frac{\partial {\cal
L}}{\partial (\partial_{\mu}\overline{\Psi})} \right)=0 .
\end{equation}
If we substitute the Lagrangian density (14) into the
Euler-Lagrange equation (15), we will obtain the Dirac equation as
follows
\begin{equation}
\left(i\hbar\gamma^{\mu}\partial_{\mu}- mc\right)\Psi=0.
\end{equation}
If we apply the Euler-Lagrange equation to $\Psi$, we obtain
\begin{equation}
i\hbar\left(\partial_{\mu}\overline{\Psi}\right)\gamma^{\mu} +
mc\overline{\Psi}=0,
\end{equation}
which is the adjoint of the Dirac equation. Now we want to obtain
the Lagrangian density for the spinor field in the presence of a
minimal length based on the Quesne-Tkachuk algebra. For such a
purpose, let us write the Lagrangian density by using the
representations (12) and (13), i.e.,
\begin{equation}
x^{\mu}\longrightarrow x^{\mu},
\end{equation}
\begin{equation}
\partial^{\mu}\longrightarrow (1+
 \beta\hbar^{2}\Box)\partial^{\mu},
\end{equation}
where $\Box := \partial_{\mu}\partial^{\mu}$ is the d'Alembertian
operator.\\
The result reads
\begin{equation}
{\cal L}= \frac{i \hbar c}{2}\left\{\overline{\Psi}
\gamma^{\mu}(\partial_{\mu}\Psi)-(\partial_{\mu}\overline{\Psi})\gamma^{\mu}\Psi
+\beta\hbar^{2} [\hspace {1mm} \overline{\Psi}\gamma^{\mu}(\Box
\partial_{\mu} \Psi)- (\Box
\partial_{\mu}\overline{\Psi})\gamma^{\mu}\Psi]\right\}-mc^{2}\overline{\Psi}\Psi
+ {\cal O} (\beta^{2}).
\end{equation}
The term $\beta \frac{i \hbar^{3} c}{2}[ \hspace {1mm}
\overline{\Psi}\gamma^{\mu}(\Box \partial_{\mu} \Psi)- (\Box
\partial_{\mu}\overline{\Psi})\gamma^{\mu}\Psi]$ in (20) can be
considered as a minimal length effect. The generalized
Euler-Lagrange equation for the adjoint spinor $\overline{\Psi}$
is [21,22]
\begin{equation}
\frac{\partial{\cal L}}{\partial
\overline{\Psi}}-\partial_{\mu}\left(\frac{\partial{\cal
L}}{\partial(\partial_{\mu}\overline{\Psi})}\right)+
\partial_{\mu}\partial_{\nu}\left(\frac{\partial {\cal
L}}{\partial(\partial_{\mu}\partial_{\nu}\overline{\Psi})}\right)
-\partial_{\mu}\partial_{\nu}\partial_{\lambda}\left(\frac{\partial
{\cal
L}}{\partial(\partial_{\mu}\partial_{\nu}\partial_{\lambda}\overline{\Psi})}\right)+\cdot
\cdot \cdot =0.
\end{equation}

If we substitute the Lagrangian density (20) into the generalized
Euler-Lagrange equation (21) and neglecting terms of order
$\beta^{2}$, we will obtain the modified Dirac equation as follows
\begin{equation}
[i\hbar \gamma^{\mu} (1+ \beta \hbar^{2}
\Box)\partial_{\mu}-mc]{\Psi}=0.
\end{equation}
The term $i\hbar^{3}\beta \gamma^{\mu}\Box
\partial_{\mu}\Psi$ in (22) shows the minimal length
effects. The wave equation (22) is a third order relativistic
wave equation that in the limit of $\beta\longrightarrow 0$ turns
in to the ordinary Dirac equation. Applying the generalized
Euler-Lagrange equation to $\Psi$, i.e.,
\begin{equation}
\frac{\partial{\cal L}}{\partial
\Psi}-\partial_{\mu}\left(\frac{\partial{\cal
L}}{\partial(\partial_{\mu}\Psi)}\right)+
\partial_{\mu}\partial_{\nu}\left(\frac{\partial {\cal
L}}{\partial(\partial_{\mu}\partial_{\nu}\Psi)}\right)
-\partial_{\mu}\partial_{\nu}\partial_{\lambda}\left(\frac{\partial
{\cal
L}}{\partial(\partial_{\mu}\partial_{\nu}\partial_{\lambda}\Psi)}\right)+\cdot
\cdot \cdot =0,
\end{equation}
and neglecting terms of order $\beta^{2}$, we find
\begin{equation}
i\hbar
[(1+\beta\hbar^{2}\Box)\partial_{\mu}\overline{\Psi}]\gamma^{\mu}+mc\overline{\Psi}=0,
\end{equation}
which is the adjoint of the modified Dirac equation.
\section{Plane-Wave Solutions of the Modified Dirac Equation}
In this section, the notation and conventions are the same as in
Greiner, Relativistic Quantum Mechanics: Wave Equations, 3rd edn
(Springer 2000) [23]. Now, we will obtain the plane-wave solutions
of the modified Dirac equation (22). The modified Dirac equation
(22) can be written as
\begin{equation}
\left[i\hbar(1+\beta\hbar^{2}\Box)\frac{\partial}{\partial t}+i\hbar
c(1+\beta\hbar^{2}\Box)\boldsymbol{\hat{\alpha}}.\boldsymbol{\nabla}-mc^{2}\hat{\beta}\right]\Psi=0,
\end{equation}
where
\begin{equation}
\hat{\beta}=\gamma^{0}= \left( {\begin{array}{*{20}c}
   I & 0  \\
   0 & -I  \\
    \end{array} } \right),\qquad
    \gamma^{i}=
    \left( {\begin{array}{*{20}c}
   0 & \sigma_{i}  \\
   -\sigma_{i} & 0  \\
    \end{array} } \right),\qquad
    \hat{\alpha}^{i}=\hat{\beta}\gamma^{i}.
\end{equation}
In equation (26) $\sigma_{i}$ are the $2\times2$ Pauli matrices, $I$
is the $2\times2$ unit matrix, and $0$ is the $2\times2$ null
matrix.\\
To solve equation (25), we try the following ansatz
\begin{equation}
\Psi(\textbf{r},t)=\psi(\textbf{r})exp(-\frac{i}{\hbar}\varepsilon
t),
\end{equation}
where $\varepsilon$ describes the time evolution of the stationary
state $\psi(\textbf{r})$. If we substitute (27) into (25), we will
obtain
\begin{equation}
[(1-\beta
\frac{\varepsilon^{2}}{c^{2}}-\beta\hbar^{2}\nabla^{2})(\varepsilon
+i\hbar c \boldsymbol{\hat{\alpha}}.\boldsymbol{\nabla})-m
c^{2}\hat{\beta}]\psi(\textbf{r})=0.
\end{equation}
The four-component spinor $\psi(\textbf{r})$ splits up into two
two-component spinors $\phi$ and $\chi$, i.e.,
\begin{equation}
\psi=\left( {\begin{array}{*{20}c}
   \psi_{1}  \\
   \psi_{2}  \\
\psi_{3}  \\
\psi_{4}
    \end{array} } \right)=\left( {\begin{array}{*{20}c}
   \phi  \\
   \chi  \\
    \end{array} } \right),
\end{equation}
with
\begin{equation}
\phi=\left( {\begin{array}{*{20}c}
   \psi_{1}  \\
   \psi_{2}  \\
    \end{array} } \right)\qquad , \qquad \chi=\left( {\begin{array}{*{20}c}
   \psi_{3}  \\
   \psi_{4}  \\
    \end{array} } \right).
\end{equation}
Using the explicit form (26) for the $\hat{\alpha}^{i}$ and
$\hat{\beta}$ matrices (28) can be written as
\begin{equation}
(1-\beta\frac{\varepsilon^{2}}{c^{2}}-\beta
\hbar^{2}\nabla^{2})\varepsilon\phi=(1-\beta\frac{\varepsilon^{2}}{c^{2}}-\beta
\hbar^{2}\nabla^{2})c\boldsymbol{\sigma}.\frac{\hbar}{i}\boldsymbol{\nabla}\chi
+mc^{2}\phi,
\end{equation}
\begin{equation}
(1-\beta\frac{\varepsilon^{2}}{c^{2}}-\beta
\hbar^{2}\nabla^{2})\varepsilon\chi=(1-\beta\frac{\varepsilon^{2}}{c^{2}}-\beta
\hbar^{2}\nabla^{2})c\boldsymbol{\sigma}.\frac{\hbar}{i}\boldsymbol{\nabla}\phi
-mc^{2}\chi.
\end{equation}
If we substitute the following ansatz
\begin{equation}
\left( {\begin{array}{*{20}c}
   \phi  \\
   \chi  \\
    \end{array} } \right)=\left( {\begin{array}{*{20}c}
   \phi_{0}  \\
   \chi_{0}  \\
    \end{array} } \right) exp(\frac{i}{\hbar}\textbf{p}.\textbf{r})
\end{equation}
into the equations (31) and (32), we will obtain
\begin{equation}
[\varepsilon(1-\beta\frac{\varepsilon^{2}}{c^{2}}+\beta\textbf{p}^{2})-mc^{2}]\phi_{0}
-c(1-\beta\frac{\varepsilon^{2}}{c^{2}}+\beta\textbf{p}^{2})(\boldsymbol{\sigma}.\textbf{p})\chi_{0}=0,
\end{equation}
\begin{equation}
-c(1-\beta\frac{\varepsilon^{2}}{c^{2}}+\beta\textbf{p}^{2})(\boldsymbol{\sigma}.\textbf{p})\phi_{0}+
[\varepsilon(1-\beta\frac{\varepsilon^{2}}{c^{2}}+\beta\textbf{p}^{2})+mc^{2}]\chi_{0}=0.
\end{equation}
So we have a linear homogeneous system of equations for $\phi_{0}$
and $\chi_{0}$, and it has nontrivial solutions only in the case
of a vanishing determinant of the coefficients, that is
\begin{equation}
\left|
\begin{matrix}
[\varepsilon(1-\beta\frac{\varepsilon^{2}}{c^{2}}+\beta\textbf{p}^{2})-mc^{2}]I &
 -c(1-\beta\frac{\varepsilon^{2}}{c^{2}}+\beta\textbf{p}^{2})(\boldsymbol{\sigma}.\textbf{p}) \\
-c(1-\beta\frac{\varepsilon^{2}}{c^{2}}+\beta\textbf{p}^{2})(\boldsymbol{\sigma}.\textbf{p})
&
[\varepsilon(1-\beta\frac{\varepsilon^{2}}{c^{2}}+\beta\textbf{p}^{2})+mc^{2}]I
\end{matrix}\right|=0.
\end{equation}
Using the identity
\begin{equation}
(\boldsymbol{\sigma}.\textbf{A})(\boldsymbol{\sigma}.\textbf{B})=\textbf{A}.\textbf{B}I+i\boldsymbol{\sigma}.(\textbf{A}\times\textbf{B}),
\end{equation}
equation (36) transforms into
\begin{equation}
(\varepsilon^{2}-c^{2}\textbf{p}^{2})(1-\beta\frac{\varepsilon^{2}}{c^{2}}+\beta\textbf{p}^{2})^{2}-m^{2}c^{4}=0.
\end{equation}
We observe that for $\beta\longrightarrow 0$, equation (38) leads to
the conventional result
\begin{equation}
\varepsilon^{2}=c^{2}\textbf{p}^{2}+m^{2}c^{4},
\end{equation}
from which follows
\begin{equation}
\varepsilon=\pm E_{p}\qquad\qquad,\qquad\qquad
E_{p}=c\sqrt{\textbf{p}^{2}+m^{2}c^{2}},
\end{equation}
as it should be. The two signs of the time evolution factor
$\varepsilon$ in (40) correspond to positive and negative energy
solutions of the conventional Dirac equation, respectively. But for
the case $\beta\neq 0$, neglecting terms of order $\beta^{2}$
provides us with two sets of results
\begin{equation}
\varepsilon_{-}=\pm E_{p}^{(-)}\qquad\qquad,\qquad\qquad
E_{p}^{(-)}=c\sqrt{\textbf{p}^{2}+ m_{-}^{2}c^{2}},
\end{equation}
\begin{equation}
\varepsilon_{+}=\pm E_{p}^{(+)}\qquad\qquad,\qquad\qquad
E_{p}^{(+)}=c\sqrt{\textbf{p}^{2}+ m_{+}^{2}c^{2}},
\end{equation}
where the non-degenerate effective masses $m_{-}$ and $m_{+}$ are
defined as
\begin{equation}
m_{-}=\frac{1}{2\sqrt{2\beta}c}\left[\sqrt{1+2\sqrt{2\beta}mc}-\sqrt{1-2\sqrt{2\beta}mc}\:\right],
\end{equation}
\begin{equation}
m_{+}=\frac{1}{2\sqrt{2\beta}c}\left[\sqrt{1+2\sqrt{2\beta}mc}+\sqrt{1-2\sqrt{2\beta}mc}\:\right].
\end{equation}
From the standpoint of quantum mechanics, (43) and (44) indicate
that our modified spinor field is associated with particles having
the effective masses $m_{-}$ and $m_{+}$. To avoid particles of
complex mass, (43) and (44) require that
\begin{equation}
\beta<\frac{1}{8m^{2}c^{2}}.
\end{equation}
It should be noted that at $\beta = \frac{1}{8m^{2}c^{2}}$ both
effective masses are equal, i.e., $m_{-}=m_{+}=m\sqrt{2}$. From
equation (35) we obtain
\begin{equation}
\chi_{0}=\frac{c(1-\beta\frac{\varepsilon^{2}}{c^{2}}+\beta\textbf{p}^{2})(\boldsymbol{\sigma}.\textbf{p})}
{\varepsilon(1-\beta\frac{\varepsilon^{2}}{c^{2}}+\beta\textbf{p}^{2})+mc^{2}}\phi_{0}.
\end{equation}
If we denote the two-component spinor $\phi_{0}$ in the form
\begin{equation}
\phi_{0}=U=\left( {\begin{array}{*{20}c}
   U_{1}  \\
   U_{2}  \\
    \end{array} } \right),
\end{equation}
with the normalization condition
$U^{\dag}U=U_{1}^{*}U_{1}+U_{2}^{*}U_{2}=1$ and using (27), (29),
(33) and (46), we will obtain two complete sets of positive and
negative energy solutions of the modified Dirac equation as
\begin{equation}
\Psi_{\textbf{p}\lambda}^{(\mp)}(\textbf{r},t)=\frac{N^{(\mp)}}{(2\pi\hbar)^{\frac{3}{2}}}\left(
{\begin{array}{*{20}c}
   U  \\
   \frac{c(1-\frac{\beta}{c^{2}}E_{p}^{(\mp)2}+\beta\textbf{p}^{2})(\boldsymbol{\sigma}.\textbf{p})}
   {\lambda E_{p}^{(\mp)}(1-\frac{\beta}{c^{2}}E_{p}^{(\mp)2}+\beta\textbf{p}^{2})+mc^{2}}U  \\
    \end{array} } \right)
    exp(\frac{i}{\hbar}(\textbf{p}.\textbf{r}-\lambda
    E_{p}^{(\mp)}t)).
\end{equation}
Here $\lambda=\pm1$ characterizes the positive and negative energy
solutions with the time evolution factors $\varepsilon_{\mp}=\lambda
E_{p}^{(\mp)}$. The normalization factors $N^{(\mp)}$ in (48) are
determined from the conditions
\begin{equation}
\int
\Psi_{\textbf{p}\lambda}^{(\mp)\dag}(\textbf{r},t)\Psi_{\textbf{p}'\lambda'}^{(\mp)}(\textbf{r},t)
d^{3}r=\delta_{\lambda\lambda'}\delta(\textbf{p}-\textbf{p}').
\end{equation}
Using equations (48) and (49) together with identity (37), the
normalization factors $N^{(\mp)}$ will be determined as
\begin{equation}
N^{(\mp)}=\left\{\frac{\left[\lambda E_{p}^{(\mp)}(1-\beta
m_{\mp}^{2}c^{2})+mc^{2} \right]^{2}}{\left[\lambda
E_{p}^{(\mp)}(1-\beta m_{\mp}^{2}c^{2})+mc^{2}
\right]^{2}+c^{2}\textbf{p}^{2}(1-\beta m_{\mp}^{2}c^{2})^{2}}
\right\}^{\frac{1}{2}}.
\end{equation}
If we expand the mass parameter $m_{-}$ in (43) to first order in
$\beta$ we will obtain
\begin{equation}
m_{-}=m+\beta m^{3}c^{2}.
\end{equation}
Inserting (51) into (41) we find the following relation
\begin{equation}
E_{p}^{(-)2}=m^{2}c^{4}+c^{2}\textbf{p}.\textbf{p}+2\beta
m^{4}c^{6},
\end{equation}
which is a modification of Einstein relation for a free particle in
special relativity. After simplification, the generalized Dirac
spinor $\Psi_{\textbf{p}\lambda}^{(-)}(\textbf{r},t)$ in (48) to
first order in $\beta$ can be written as
\begin{equation}
\Psi_{\textbf{p}\lambda}^{(-)}(\textbf{r},t)
=\frac{1}{(2\pi\hbar)^{\frac{3}{2}}}\sqrt{\frac{\lambda
E_{p}^{(-)}(1-\beta m^{2}c^{2})+mc^{2}}{2\lambda E_{p}^{(-)}(1-\beta
m^{2}c^{2})}} \left( {\begin{array}{*{20}c}
   U  \\
   \frac{c(1-\beta m^{2}c^{2})(\boldsymbol{\sigma}.\textbf{p})}{\lambda E_{p}^{(-)}(1-\beta m^{2}c^{2})+mc^{2}}U  \\
    \end{array} } \right)
    exp(\frac{i}{\hbar}(\textbf{p}.\textbf{r}-\lambda
    E_{p}^{(-)}t)),
\end{equation}
where $E_{p}^{(-)}$ was given in (52).\\
It is clear that for $\beta\longrightarrow 0$ the generalized
Dirac spinor $\Psi_{\textbf{p}\lambda}^{(-)}(\textbf{r},t)$ in
(53) will be converted into the conventional Dirac spinor
$\Psi_{\textbf{p}\lambda}(\textbf{r},t)$ for a free particle in
relativistic quantum mechanics [23], i.e.,
\begin{equation}
\Psi_{\textbf{p}\lambda}(\textbf{r},t)=\lim_{\beta\rightarrow
0}\Psi_{\textbf{p}\lambda}^{(-)}(\textbf{r},t)=\frac{1}{(2\pi\hbar)^{\frac{3}{2}}}\sqrt{\frac{\lambda
E_{p}+mc^{2}}{2\lambda E_{p}}}\left( {\begin{array}{*{20}c}
   U  \\
   \frac{c(\boldsymbol{\sigma}.\textbf{p})}{\lambda E_{p}+ mc^{2}}U  \\
    \end{array} } \right)
    exp(\frac{i}{\hbar}(\textbf{p}.\textbf{r}-\lambda E_{p}t)).
\end{equation}
For small $\beta$ the effective mass $m_{+}$ in (44) reduces to
\begin{equation}
m_{+}=\frac{1}{\sqrt{2\beta}c}-\frac{m^{2}}{2}\sqrt{2\beta}c,
\end{equation}
which diverges for $\beta\longrightarrow 0$.\\
Thus we have two massive particles in our theory, one with the
usual mass $m$ $(\lim_{\beta \rightarrow 0}\: m_{-})$ and the
other a heavy-mass particle of mass
$\frac{1}{\sqrt{2\beta}c}(\lim_{\beta\rightarrow 0}\: m_{+})$
which, leads to an indefinite metric in our model. Until all the
energies of the system are kept below the production threshold of
the $\frac{1}{\sqrt{2\beta}c}-$mass particle, the indefinite
metric does not enter and the theory obeys all physical
requirements such as unitarity. The generalized Dirac spinor
$\Psi_{\textbf{p}\lambda}^{(+)}(\textbf{r},t)$ in (48) which
describes a particle with effective mass $m_{+}$ is entirely new
and does not have a counterpart in the conventional Dirac
equation. The essential feature that is responsible for the
existence of the generalized Dirac spinor
$\Psi_{\textbf{p}\lambda}^{(+)}(\textbf{r},t)$ is that the
modified Dirac equation (22) is third order in space-time
derivatives in the presence of a minimal length, whereas the
ordinary Dirac equation is
only first order in space-time derivatives.\\
Now, let us estimate the numerical value of the minimal length in
our work. By putting $\beta'=2\beta$ into (6) and neglecting terms
of order $\beta^{2}$, the isotropic minimal length becomes
$(\bigtriangleup X^{i})_{0}\simeq\hbar \sqrt{5\beta}$. The upper
bound for deformation parameter $\beta$
$(\beta_{upper-bound}\simeq\frac{1}{8m^{2}c^{2}})$ together with
isotropic minimal length $\hbar\sqrt{5\beta_{upper-bound}}$ for an
electron $(m_{e}=9.11\times 10^{-31}kg)$ are respectively
$\beta_{upper-bound}\simeq1.67\times10^{42}\frac{s^{2}}{kg^{2}m^{2}}$
and $(\bigtriangleup X^{i})_{0}\simeq 3\times 10^{-13}m$.

\section{Conclusions}
Heisenberg in 1938 wrote an article about the significance of a
minimal length in physics [24]. He believed that every theory of
elementary particles must contain a minimal length besides two
fundamental constants, $c$ and $h$. The hope was that the
introduction of such a minimal length would eliminate the
ultraviolet divergencies from quantum field theory. This minimal
length scale leads to a GUP.  An immediate consequence of the GUP
is a generalization of momentum operator according to (13) for
$\beta'=2\beta$. This generalized form of momentum operator leads
to a modified Dirac equation. We have shown that our modified
Dirac equation which contains higher order derivative of the wave
function describes two massive particles, one particle with the
effective mass $m_{-}$ and the other a very heavy particle with
the effective mass $m_{+}$ according to (43) and (44). From (43)
and (44) the restriction on the deformation parameter $\beta$
becomes $\beta<\frac{1}{8m^{2}c^{2}}$. This restriction leads to
an isotropic minimal length $(\triangle
X^{i})_{0}\simeq\frac{\sqrt{10}}{4}\frac{\hbar}{mc}$. In [13-18]
considering the Lamb shift the authors estimated $(\triangle
X^{i})_{0}\leq 10^{-16}- 10^{-17} m$, analysis of electron motion
in a Penning trap also gives $(\triangle X^{i})_{0}\leq
10^{-16}m$ [19]. The obtained value for upper bound of the
isotropic minimal length in this work is $(\triangle
X^{i})_{0}\simeq 3\times 10^{-13}$m. Although the above value for
the isotropic minimal length is about 2 orders of magnitude
larger than that was proposed by Heisenberg $(10^{-15}m)$ [25],
this value is near to the reduced Compton wavelength of the
electron $(\mathchar'26\mkern-10mu\lambda_{c}=
\frac{\hbar}{m_{e}c}=3.86\times 10^{-13} m)$.

\section*{Acknowledgements}
We are grateful to M. Poorakbar for helpful discussions.

\end{document}